\documentclass[%
  reprint,
  superscriptaddress,
showpacs,preprintnumbers,
nofootinbib,
 amsmath,amssymb, 
 aps,
 prd,
floatfix,
showkeys,
]{revtex4-2}

\usepackage{amsmath,amstext,amssymb}
\usepackage[T1]{fontenc}
\usepackage{aas_macros}
\usepackage{xcolor}
\usepackage{soul}
\usepackage{comment}
\usepackage{graphicx}
\usepackage[normalem]{ulem}
\usepackage[export]{adjustbox}
\usepackage[colorlinks=true,allcolors=blue,pdfencoding=auto, psdextra]{hyperref}
\usepackage{orcidlink}
\usepackage{multirow}
\usepackage{siunitx}

\begin{document}

\title[BAO 2D with S-PLUS]{A cosmology weakly dependent measurement of 2D Baryon Acoustic Oscillations scale from the Southern Photometric Local Universe Survey}
\author{U.~Ribeiro~}\email{ulissesrs@cbpf.br}
\affiliation{Centro Brasileiro de Pesquisas F\'isicas, Rua Dr. Xavier Sigaud 150, 22290-180 Rio de Janeiro, RJ, Brazil}

\author{F.~Avila~\orcidlink{0000-0002-0562-2541}}\email{fsavila2@gmail.com}
\affiliation{Observatório Nacional, Rua General José Cristino, 77, São Cristóvão, 20921-400, Rio de Janeiro,  RJ, Brazil}

\author{C. R. Bom~\orcidlink{0000-0003-4383-2969}}\email{debom@cbpf.br}
\affiliation{Centro Brasileiro de Pesquisas F\'isicas, Rua Dr. Xavier Sigaud 150, 22290-180 Rio de Janeiro, RJ, Brazil}

\author{C.~Franco~\orcidlink{0000-0002-6320-425X}}\email{camilafranco@on.br}
\affiliation{Observatório Nacional, Rua General José Cristino, 77, São Cristóvão, 20921-400, Rio de Janeiro,  RJ, Brazil}

\author{A.~Cortesi~}\email{}
\affiliation{Instituto de F\'{i}sica, Universidade Federal do Rio de Janeiro, 21941-972, Rio de Janeiro, Brazil}\affiliation{Observat\'{o}rio do Valongo, Ladeira Pedro Antonio, 43, 20080-090, Rio de Janeiro, Brazil}

\author{C.~Mendes de Oliveira}
\affiliation{Universidade de S\~ao Paulo, IAG, Rua do Matão 1225, São
Paulo, SP, Brazil}
\author{A. Kanaan}
\affiliation{Departamento de F\'isica, \\ Universidade Federal de Santa Catarina, Florian\'opolis, SC, 88040-900, Brazil}

\author{T. Ribeiro}
\affiliation{Rubin Observatory Project Office, 950 N. Cherry Ave., Tucson, AZ 85719, USA}

\author{W. Schoenell}
\affiliation{GMTO Corporation 465 N. Halstead Street, Suite 250 Pasadena, CA 91107, USA}

\author{E. Telles}
\affiliation{Observatório Nacional, Rua General José Cristino, 77, São Cristóvão, 20921-400, Rio de Janeiro,  RJ, Brazil}

\author{A.~Bernui~\orcidlink{0000-0003-3034-0762}}\email{bernui@on.br}
\affiliation{Observatório Nacional, Rua General José Cristino, 77, São Cristóvão, 20921-400, Rio de Janeiro,  RJ, Brazil}



\begin{abstract}
Baryon Acoustic Oscillations (BAO) provide a robust standard ruler for observational cosmology, enabling precise constraints on the expansion history of the Universe. 
We present a weakly model-dependent measurement of the BAO angular scale in the low-redshift Universe using the blue galaxies from the Southern Photometric Local Universe Survey (S-PLUS). Our analysis is based on the 2-point angular correlation function applied to a selected photometric sample of $5977$ galaxies with redshifts $0.03 \leq z \leq 0.1$. To account for photometric redshift uncertainties, we implement a resampling technique using the probability distribution function of each galaxy. Angular correlations are computed using the Landy-Szalay estimator; the uncertainties are quantified using a set of $1000$ log-normal mock catalogues. 
Our 2-point angular correlation analyses reveal a prominent BAO signal that after a shift correction, due to the projection effect caused by the finite thickness of the redshift bin, provides 
the transversal BAO measurement: $\theta_{\rm BAO} = 21.81^{\circ} \pm 0.85^{\circ}$, at $z_{\rm eff} = 0.075$, 
detected with a statistical significance of $3.22\,\sigma$. 
In addition, we performed consistency tests that support the robustness of our result. 
Our measurement constitutes the first robust detection of the transversal BAO scale: at the lowest-redshift in the Universe and using multi-band (narrow+wide) photometry data from the S-PLUS.
\end{abstract}




\maketitle
\section{Introduction}\label{sec1}

The formation of cosmic structures is driven by primordial processes 
where density fluctuations evolved under gravitational 
and electromagnetic interactions 
in an expanding universe. 
Among these, Baryon Acoustic Oscillations (BAO) phenomenon 
causes sound waves propagating in the photon-baryon plasma 
until recombination~\citep{Peebles1970,Sunyaev1970}. 
At the epoch of recombination, photons decouple from baryons, 
freezing these oscillations into the matter distribution, leaving 
as a characteristic imprint on the large-scale structure 
the size of the frozen spherical wave: the sound horizon 
at drag epoch $r_s$. 
This scale can be recovered studying the three-dimensional (3D) distribution of cosmic objects in comoving coordinates, 
where this scale remains constant at any epoch of the universe. 
For this, the BAO phenomenon is a valuable cosmological standard ruler, providing a robust means to measure the expansion history of the universe~\citep{Cole05,Eisenstein05,Bassett10,Cosmoverse}. 

Traditionally, the BAO signature has been studied using 3D statistics, like the 2-point correlation function or the power spectrum, both requiring a fiducial cosmology to convert redshifts into comoving distances~\citep{Eisenstein05,Blake11}. 
However, an alternative study consists of measuring the transversal BAO 
signature, approach that relies solely on angular correlations within 
a redshift bin, with weak dependence on cosmological models for a shift correction~\citep{Sanchez2011,carnero2011tracing, carvalho2016}. 
In fact, due to the finite size of the redshift bin the projection effect makes necessary an angular shift in the BAO signal observed in the 2-point angular correlation function~\citep{deCarvalho2018,deCarvalho2021}.

Recent studies have demonstrated the feasibility of detecting 2D BAO using photometric surveys, despite challenges such as large photometric-redshift uncertainties~\citep{ansari2019impact,crocce2019dark,DES2022,Sanchez2011}. 
Spectroscopic data provide precise redshift estimates, but they are observationally expensive, limiting the volume and number of tracers available. 
Instead, photometric surveys, on the other hand, offer larger datasets but require careful analyses due to large redshift uncertainties. 
For this reason, performing BAO analysis with photometric redshift data is, indeed, challenging \cite{chaves2018effect, ansari2019impact, ishikawa2023robustness, ferreira2025influence}.

In this work, we develop a methodology to efficiently deal with this issue. We used the Probability Distribution Function (PDF) of the photometric redshift of each cosmic object to generate a set of new photometric catalogs, as we shall describe below, to obtain a robust measurement of the transverse BAO signal in a redshift bin of the Local Universe. 
For this task, we first select the sample of blue galaxies with the highest accuracy in the redshift measurement from the recently released S-PLUS photometric survey data. 
We then measure the BAO angular scale at the effective redshift $z_{\text{eff}} = 0.075$, 
and obtain the angular diameter distance $D_A(z_{\text{eff}})$. 
We perform consistency tests along the analyses done, 
and also assess the impact of photometric uncertainties on 
the BAO signal found recalculating the transversal BAO signature 
using simulated mock catalogs. 
Additionally, we calculate the statistical significance of our transversal BAO measurement.

The structure of this work is as follows:
in Section~\ref{sec:data}, we describe the dataset and selection criteria for the galaxy sample. 
Section~\ref{sec:methodology} outlines the methodology for computing the two-point angular correlation function and mitigating systematic effects. 
Section~\ref{sec:results} presents our results, including the detection of the BAO bump and its implications for cosmology. 
Finally, we summarize our conclusions in Section~\ref{sec:conclusion}.


\section{Data} \label{sec:data}

The Southern Photometric Local Universe Survey  (S-PLUS) is an astronomical survey that maps the Local Universe with multi-band photometry with 12 filters:  7~narrow-band filters and 5~large-band filters (i.e., SDSS-type filters)~\citep{splus}. 
The telescope, located in Cerro Tololo (Chile), has a 0.86m mirror which, combined with a field of view of 1.4 square degrees and an 85 Mega-pixel camera, is producing high-quality images with a unique spectral resolution for millions of objects over several thousand square degrees.
The narrowband filters, in fact, are centred over specific absorption and emission lines, namely the
$[OII]$, Ca H + K, H$\delta$, G band, Mgb triplet, $H_{\alpha}$, and Ca triplet
features, corresponding to the J0378, J0395, J0410,
J0430, J0515, J0660, J0861 filters. The system also includes the u, g, r, i, and z broadband filters which serve to constrain the spectral continuum of sources. 
The filter system coverage render possible to estimate photometric redshift with a precision of 
$\sigma_z \simeq 0.03(1+z)$ \citep{splus}, as well as retrieve stellar population and morphological properties of the sources. This multi-purpose astrophysical survey in the southern hemisphere  will observe more than $8000$ square degrees (1/5 of the whole sky), covering the entire visible region of the electromagnetic spectrum ($3500$ \si{\angstrom} to $10000$ \si{\angstrom}). 
The latest internal data release (iDR5) covers 
$4592.2$ square degrees.

\subsection{S-PLUS blue galaxies data}
\label{subsec:blue galaxies}

The Local Universe is full of galaxies showing diverse 
morphologies, colour, and environments, i.e., a true galaxy zoo~\citep{lintott2008,Bamford2009}. 
As large astronomical surveys emerged, detailed examinations regarding 
the galaxy clustering dependence on colour and luminosity, particularly in samples of red and blue galaxies, has been 
reported~\citep{Zehavi05, Croton07, Ross14, 
Mohammad18}.
Blue galaxies are, in their majority, late-type galaxies with significant star formation, meaning they are unlikely to be found in high density regions \citep{Dressler, Gerke07, Dias2023}, 
a feature that is reflected in clustering statistics as the 2-point correlation function, where, on small scales, red galaxies of any luminosity are more clustered than blue galaxies of any luminosity~\citep{Zehavi05}. 
Indeed, blue galaxies are found in low-density regions where they exhibit reduced non-linear clustering effects, making the 
sample of blue galaxies the suitable cosmic tracer for BAO analyses~\citep{Gerke07, Mohammad18, deCarvalho2021, Avila24}. 
Moreover, blue galaxies account for nearly $60\%$ 
of galaxies in the Local 
Universe~\citep{Bamford2009}, being, therefore, 
also more numerous in relation to their red 
counterparts. 
 

Our first task is, therefore, the careful choice of the sample of blue galaxies for our BAO analyses. 
For this process, we use the observed parameters released within the fifth internal 
data release (iDR5) of the S-PLUS. 

Specifically, we use the star-galaxy probability, $PROB\_GAL\_GAIA$ 
and $CLASS\_GAIA$ as defined in \cite{nakazono2021discovery}, and the error on the photo z estimations, $odds$ ~\citep{lima2022photometric}. 
The other parameters are the standard outputs of Sextractor code, 
which are also employed in the extraction of the catalogs (see the documentation in S-SPLUS Cloud\footnote{https://splus.cloud/documentation}). 
In brief, the selection criteria are as follows: 
\begin{itemize}
\item S1: $(r\_auto >= 14)\,\textsc{and}\,(r\_auto \le 18)$
\item S2: $PROB\_GAL\_GAIA > 0.9$
\item S3: $FWHM\_n \ge 1.5$
\item S4: $SEX\_FLAGS\_r \le 2$
\item S5: $CLASS\_STAR \le 0.2$
\item S6: $CLASS\_GAIA = 2$
\item S7: $odds \ge 0.4$
\end{itemize}
The restriction in magnitude aims at excluding very large objects, which are often misclassified \citep{nakazono2021discovery} and to exclude very faint objects, which are also harder to classify. 
Moreover, we applied the K-correction to all magnitudes 
(see~\cite{bom2024extended} De Bom et al. 2024).

\begin{figure}
\centering
\includegraphics[width=1.1\linewidth]{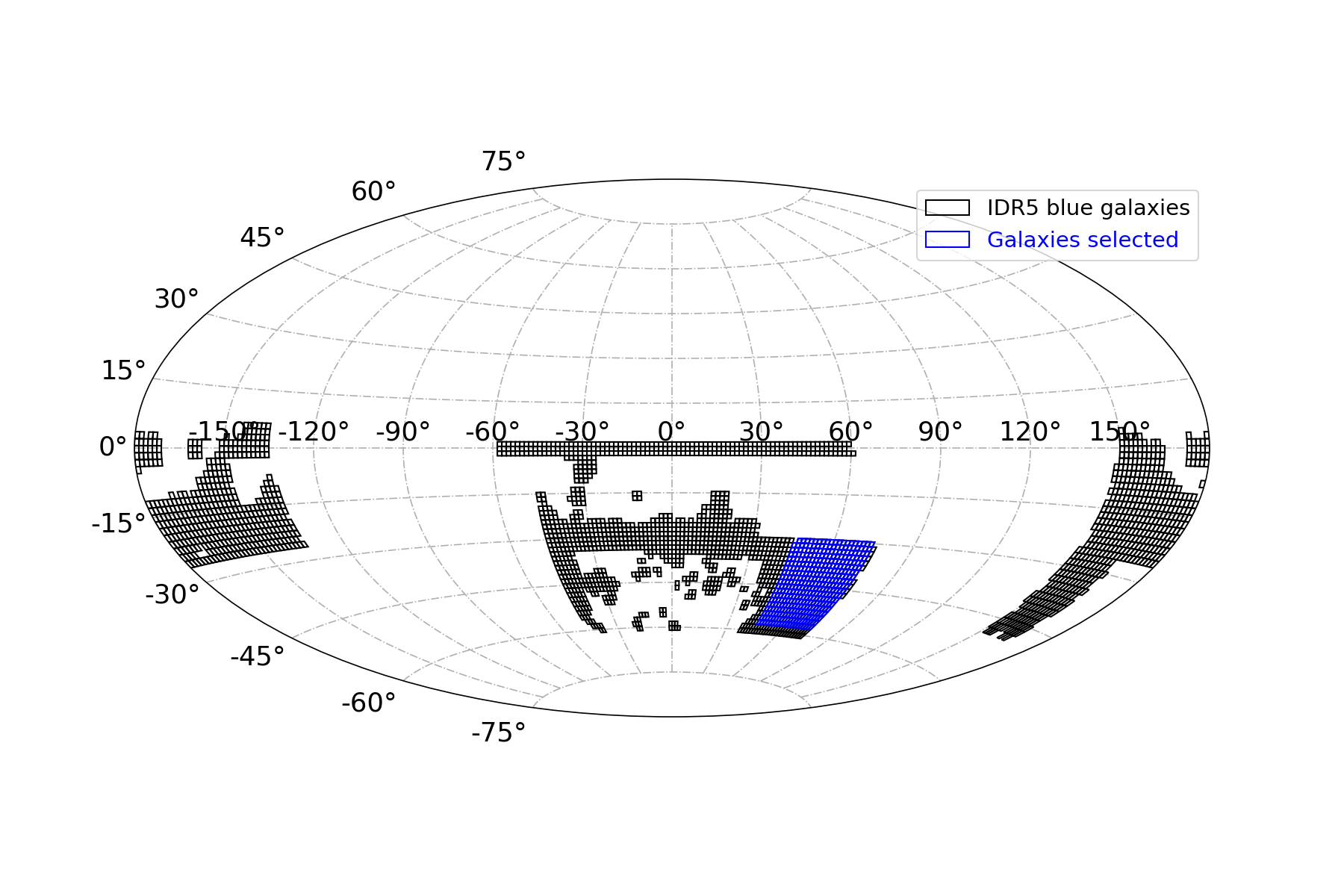}
\caption{Footprint covering all blue galaxies of S-PLUS iDR5 given the selections used in this work. The region highlighted in blue is the footprint selected for our analyses.}
\label{fig:footprint}
\end{figure}

\begin{figure}
\centering
\includegraphics[width=1\linewidth]{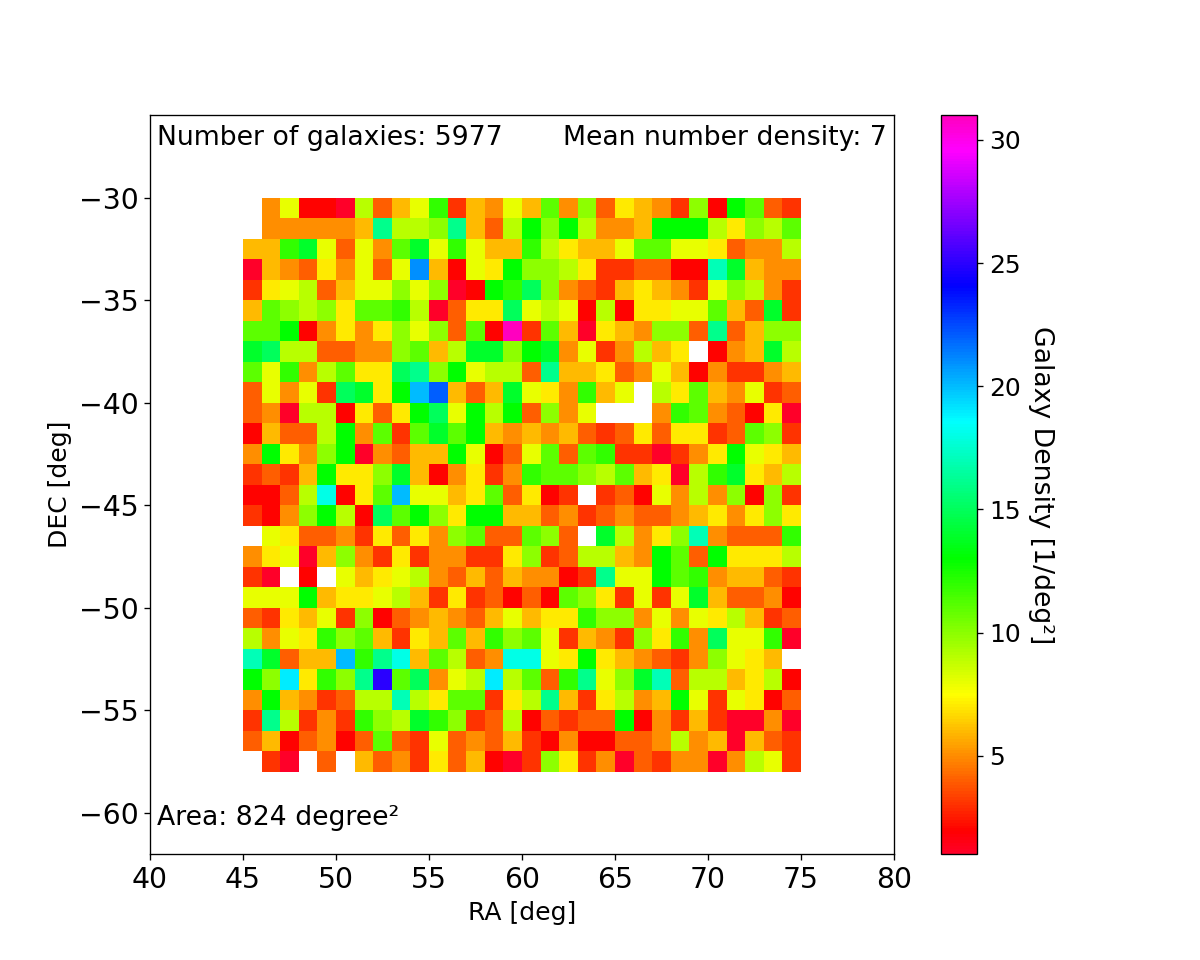}
\caption{Footprint of the selected region for our BAO analyses. Each bin has an area of 1 square degree being colored in terms of the number of galaxies that falls in it. The data from this sky region correspond to the 5th internal data 
release (iDR5) of the S-PLUS. }
\label{fig:density_map}
\end{figure}

\begin{figure}
\centering
\includegraphics[width=0.95\linewidth]{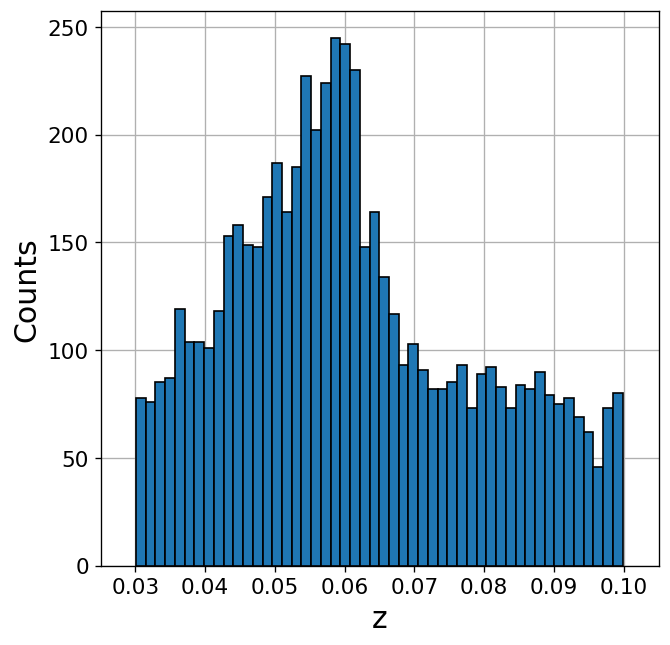}
\caption{Histogram of the data observed in the footprint shown in figure~\ref{fig:footprint}, 
corresponding to the S-PLUS iDR5. 
We have limited our sample for analysis for galaxies with redshift 
$0.03 \le z \le 0.10$; the lower limit was chosen to avoid non-linear features that are difficult to reproduce in the mock catalogs, while 
the upper limit was selected due to the decrease in the number density of galaxies for larger redshifts.
The median redshift value of this data sample is $\bar{z} = 0.058$.
} 
\label{fig:geometry}
\end{figure}

In order to select the blue galaxies from this sample we applied the following cuts in colours~\citep{Avila19} 
\begin{itemize}
\item
\textrm{S8}: 
$(0 < g - r < 0.6)\,\textsc{and}\,(0 < u - r < 2.0)$ 
\end{itemize}
and, to improve the quality of the catalog, we also excluded objects with large errors in colours. 
These dataset was then divided into four groups of contiguous regions, among which the third region was selected to perform this study, given its extension and numerical density. 
The final sample chosen for our analyses contains $N_g = 5977$ 
blue galaxies distributed in the region of right ascension RA = [45.00,74.98] and declination 
Dec = [-57.99, -30.18] (see figure~\ref{fig:footprint}). 
The region covers an area of approximately 824 square degrees with a good density of points as shown in 
figure~\ref{fig:density_map}, that is, 7.25 galaxies per square degree. 
The redshift distribution of this sample, $0.03 \le z \le 0.1$, 
is shown in figure~\ref{fig:geometry}.

\subsection{Random catalogs}

A random sample is a realization of a statistically homogeneous distribution of cosmic objects in a given volume of the universe. 
Deviations from a uniform distribution represents matter clustering, an effect that is quantified by a two-point correlation function estimator, by comparison of the data sample with a substantial number of random 
catalogs (see, e.g.,~\citep{Franco2024, deCarvalho2018, Keihanen19, Wang13}).

In this study, we generate the random catalog in the same footprint of the data. 
Accordingly, we applied the same cuts in RA and Dec as those used in the observational sample. 
That is, the random sample is generated with the same geometry and volume of the data sample, therefore the comparison between them in the angular correlations examination will be valid. 
Additionally, having the same geometry and volume helps mitigating border effects. 
Outliers in the data sample will also be outliers in the random sample, which keeps the galaxy counting consistent.

To suppress statistical noise in the correlation function 
--particularly, in the random-random and data-random pair counts-- we adopted a random sample containing 10 times more objects than the data. 
This choice improves the statistical robustness of the correlation function measurement by reducing the shot-noise contribution associated with the random catalog (see, e.g.,~\cite{Keihanen19,Avila24,Franco25}).

\subsection{Simulated catalogs (mocks)}
\label{sec:mocks}

The statistical significance of BAO detection relies critically on a robust estimation of the uncertainties associated with the angular correlation measurements~\citep{Eisenstein05,Cole05,Beutler11,Carter18}. In this work, we adopt pseudo-simulations known as log-normal realizations~\citep{Coles91}. Previous studies have demonstrated that log-normal simulations provide a reliable framework for estimating uncertainties in large-scale structure analyses~\citep{Lippich19,Blot19,Colavincenzo19}.

The underlying idea is to transform a Gaussian field $X(r)$ into a positive-definite field $Y(r)$ via the exponential mapping $Y(r)=\exp[X(r)]$, yielding a log-normal distribution. This approach overcomes the limitations of Gaussian models, such as the possibility of assigning non-zero probability to negative densities.

We use the public code\footnote{\url{https://bitbucket.org/komatsu5147/lognormal_galaxies/src/master/}} developed in \citep{Agrawal17}. To create the simulations, we input cosmological parameters, the survey settings and a non-linear power spectrum, the latter generated with the CCL code\footnote{\url{https://github.com/LSSTDESC/CCL}}~\citep{Chisari19}. The table \ref{table1} summarizes the input values for generating the log-normal simulated catalogs 
or mocks. 
With this procedure we generated $N_m=1000$ mock catalogs 
for our analyses.

\begin{table}[h]
\caption{Survey configuration and cosmological parameters used to generate the set of $N_m = 1000$ mock catalogs. 
$N_{g/m}$ is the number of galaxies for this mocks production.}
\centering
\begin{tabular}{c|c}
	\hline
	Survey configuration          & Cosmological parameters       \\ 
\hline
	$z=0.075$                     & $\Omega_{c}h^{2}= 0.1202$     \\  
	$b=1.0$                       & $\Sigma m_{\nu}=0.06$         \\
	$N_{g/m}=6\,000\,000$     & $n_{s}=0.9649$                    \\
	$L_{x}=340\ \textrm{Mpc/$h$}$                     & $\ln(10 A_{s})=3.045$                  \\
	$L_{y}=431\ \textrm{Mpc/$h$}$                     & $\Omega_{b}h^{2}=0.02236$       \\
	$L_{z}=418\ \textrm{Mpc/$h$}$                     & $h=0.67021$       \\ \hline                          
\end{tabular}
\label{table1}
\end{table}

\section{Methodology} \label{sec:methodology}

In this section we describe our approach to search for a statistically significant transversal BAO signature in the selected sample of blue galaxies from the iDR5 of S-PLUS. 
Firstly, we shall explain the theoretical framework to calculate the 2-point correlation function and then its angular version, the 2-point angular correlation function. 
After that we describe how to compute the covariance matrix that provides the uncertainties in the analysis of the angular correlations. 
Then, we study how to find the angular scale of the transversal BAO signature, that appears as a bump\footnote{This bump is the manifestation of the excess probability of finding a set of blue galaxies forming rings, or parts of them, on the sky.} in the correlation function analysis at the angular scale 
$\theta_{\rm FIT}$, 
which is then corrected for the projection effect to get the angular BAO scale, $\theta_{\rm BAO}$. 
Finally, we determine the statistical significance of our transversal BAO measurement.

\subsection{Correlation Function} \label{subsec:2PCF}
\subsubsection{Spatial Correlation Function}

The 2-point correlation function (2PCF) is a statistical tool that let us to extract useful information about 
the matter distribution in the universe~\citep{Avila22, Marques20}. 
The direct interpretation for the 2PCF is that it determines an excess or missing probability of finding two point sources with a distance $r$ and can be derived from the matter power spectrum with the equation~\cite{Sanchez2011} 
\begin{equation}
\xi(r;z) = \int_0^\infty \frac{dk}{2\pi^2}k^2j_0(kr)P(k;z) \,,
\end{equation}
where $j_0$ is the zeroth-order Bessel function and $P(k;z)$ is the matter power spectrum. Naturally, one can use codes such as \textsc{CAMB} and \textsc{CLASS}, or use simulations to compute the matter power spectrum and then obtain the 2PCF. 
Another thing to take into consideration is that the power spectrum is associated with the matter energy content of the universe. Usually, a linear bias $b$ is introduced to translate from the matter density field to the tracer used, that is, the S-PLUS blue galaxies, as
\begin{equation}
\xi_{gg}(r;z) = b(z)^2 \,\xi(r;z)\,.
\end{equation}
\subsubsection{Angular correlation Function}

The 2-point angular correlation function (2PACF), 
$\omega(\theta)$, study the angular separation between pairs of cosmic objects in a given redshift bin. It can be derived in a similar approach as the one used in the spatial correlation function, but with the data projected in the celestial sphere. Therefore, we use the angular power spectrum $C^{ab}_l$ that can be calculated for two tracers $a$ and $b$ with the equation

\begin{equation}
\label{eq:angular power spectrum}
    C^{ab}_\ell = 4\pi \int \frac{dk}{k}\mathcal{P}_\Phi(k)\Delta^a_\ell(k)\Delta^b_\ell(k),
\end{equation}
where $\mathcal{P}_\Phi(k)$ is the dimensionless power spectrum of the primordial curvature perturbations, with $\Delta^a_\ell(k)$ and $\Delta^b_\ell(k)$ being the transfer functions to these tracers. For the case of discrete sources the transfer function based on number counts can be calculated as

\begin{equation}
\label{eq:tranfer function}
    \Delta_\ell(k) = \int dz\ p_z(z)b(z)T_\delta(k,z)j_\ell(k\chi(z)),
\end{equation}
where $p_z(z)$ is the normalized distribution of sources in redshift, $T_\delta$ is the matter overdensity transfer function and $j_\ell(x)$ is the $\ell$-th order Bessel function. Having the angular power spectrum one can obtain the angular correlation function with the equation 
\begin{equation}
\label{eq:2PACF theory}
w(\theta) = \sum_{\ell} \frac{2\ell + 1}{4\pi} \, C_\ell^{ab} \, P_{\ell}(\cos\theta),
\end{equation}
where $P_{\ell}$ are the Legendre polynomials of order $\ell$. 
These calculations were performed with the public code CCL\footnote{https://github.com/LSSTDESC/CCL}.

%
%

\subsection{Correlation Function Estimator} \label{subsec:correlation function estimator}

%

There are several estimators to calculate the 2PACF, $\omega(\theta)$, used to calculate matter clustering in the data by counting the number of pairs of cosmic objects with angular distance $\theta$ in the data sample and then comparing with a similar count in a random sample. 
In our analyses, we use the Landy-Szalay estimator~\citep{landy1993bias}, 
which reduces its bias and variance on large scales, 
$\omega(\theta) \ll 1$, 
%
\begin{equation} \label{eq:2PACF estimator}
\omega(\theta) = \frac{dd(\theta) - 2dr(\theta) + rr(\theta)}{rr(\theta)} \,, 
\end{equation}
where  
\begin{align*}
dd(\theta) &\equiv \frac{DD(\theta)}{n_g\,(n_g-1)/2} \,,\\
\notag\\
rr(\theta) &\equiv \frac{RR(\theta)}{n_R\,(n_R-1)/2} \,,\\
\notag\\
dr(\theta) &\equiv \frac{DR(\theta)}{n_g\,n_R}\,,
\end{align*}
with $n_g$ ($n_R$) being the number of galaxies in the data (random) sample in analysis, and 
where the quantities $DD$, $DR$, and $RR$ were normalized by the number of pairs (to provide the estimator with an interpretation in terms of probabilities), and are defined as 
\begin{itemize}
\item $DD(\theta)$ is the number of pairs of galaxies with angular distance $\theta$ from each other with $\delta \theta$ of tolerance. The pairs are counted considering galaxies in the data. 
\item $RR(\theta)$ is the number of pairs of galaxies with angular distance $\theta$ from each other with $\delta \theta$ of tolerance. The pairs are counted considering galaxies in the random. 
\item $DR(\theta)$ is the number of pairs of galaxies with angular distance $\theta$ from each other with $\delta \theta$ of tolerance. The pairs are counted considering one galaxy in the data and one galaxy in the random. 
\end{itemize}



These calculations were done using the code \textsc{TreeCorr}\footnote{https://rmjarvis.github.io/TreeCorr} \citep{jarvis2004skewness} that incorporates those estimators for the correlation function in its package.

\subsection{Covariance Matrix}\label{covariance-matrix}

\begin{figure}[ht]
\centering
\includegraphics[width=0.8\linewidth]{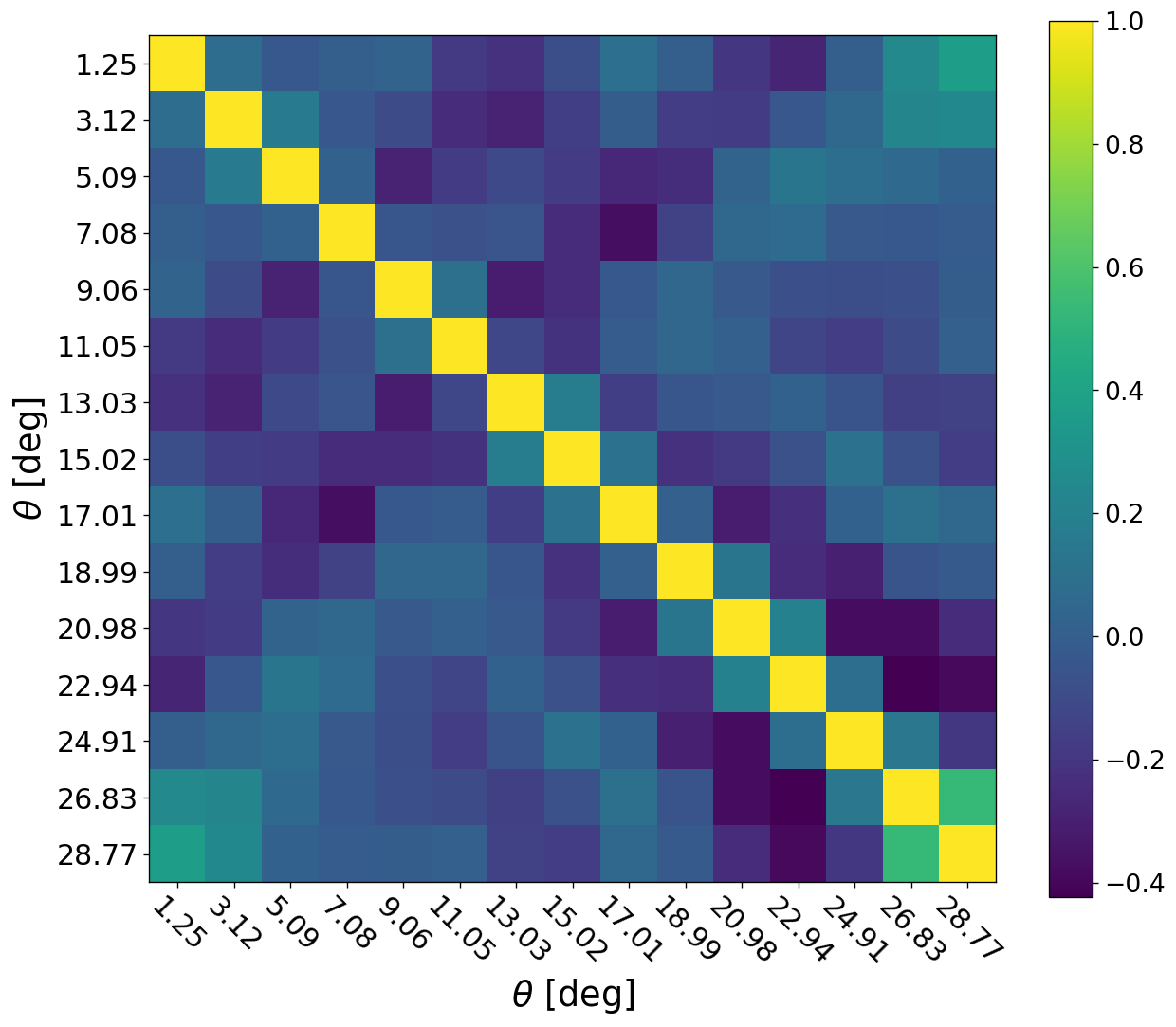}
\caption{Correlation matrix calculated with the $N_m=1000$ 
mock catalogs, described in section~\ref{sec:mocks}. }
\label{fig:correlation-matrix}
\end{figure}

The covariance matrix was calculated with 1000 mock catalogs created as described in Section~\ref{sec:mocks}. Similarly to the photometric error, we calculate the 2PACF for each mock and compute the covariance matrix for each discrete angular separation with the following formula,

\begin{equation}
    \textrm{Cov}_{ij} = \frac{1}{N}\sum_k^N \left[ \omega_k(\theta_i) -  \overline{\omega_k}(\theta_i) \right] \left[ \omega_k(\theta_j) -  \overline{\omega_k}(\theta_j) \right],
\end{equation}
where $\theta_i$ is the i-th angular separation calculated using \textsc{TreeCorr}, $\omega_k(\theta_i)$ is the 2PACF calculated in the k-th mock and $\overline{\omega_k}(\theta_i)$ is the average over all mocks. With this, we can estimate the angular correlation function error for each angular separation by taking the square root of the elements in the main diagonal (see figure \ref{fig:correlation-matrix}). 

\subsection{Sampling the data catalog: Photometric Error}
\label{sampling-catalogs}

Although the redshift is not directly used to calculate the angular correlation function we still need it to decide in which redshift bin each galaxy falls. 
Naturally we need to have a redshift bin $\Delta z$ larger than the average redshift uncertainty $\langle \sigma_z \rangle$ to contemplate possible deviations. This systematic error can change the number of pairs in the bin, making cosmic objects to appear or disappear from our selected sample, and a possible source of bias in our BAO analysis. In fact, this effect is reduced when $\langle \sigma_z \rangle $ is low when compared to $\Delta z$, 
$$\frac{\langle\sigma_z\rangle}{\Delta z} \ll 1 \,,$$
because, mainly, galaxies close to the border would present this problem. 
In the present case, the width of the redshift bin in analysis is 
$\Delta z = 0.07$, and $\langle \sigma_z \rangle = 0.014$ which is $20\%$ of the redshift bin width. 
This value is indeed large if compared with the case with 
spectroscopic redshifts where this quotient will reach values 
of about $1\%$. 

The techniques used to determine redshift using photometry 
tends to have higher uncertainties than those with spectroscopy, but it is not a good solution to keep increasing the bin width to contemplate those deviations, once it would destroy the signal. One way to think about it is: with higher uncertainties, we have more possible configurations for the pairs of galaxies inside the bin. 
For each of those configurations, we calculate our 2PACF and look for the BAO signal. 
Each configuration can be taken as a fluctuation of the data and the photometric error can be introduced from them.

S-PLUS provides the photo-z PDF for each of the galaxies, so the method can be described as follows:
 \begin{itemize}
     \item Use the PDF to construct the Cumulative Probability Function 
     for each galaxy and implement the sampling technique known as inverse transform sampling. With this, new redshifts are randomly chosen following the PDF of their respective galaxies.

     \item Construct the fluctuations of the data taking the redshift samples and separating the galaxies in the desired redshift bins. 

     \item Calculate the 2PACF with Equation~\eqref{eq:2PACF estimator} for each configuration and take the average over all correlation functions for every discrete angular separation.

     \item If the signal persists in the end of this process, apply the parametric model of Equation~\eqref{eq:parametric model} to it.
 \end{itemize}

\begin{figure}
\centering
\includegraphics[width=1\linewidth]{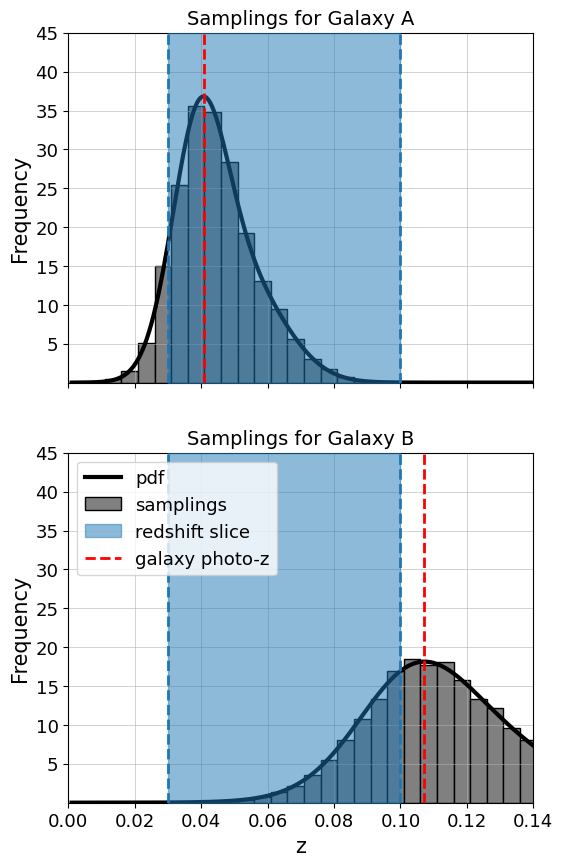}
\caption{Samplings for a galaxy redshift. 
In the upper panel we show the sampling of a galaxy redshift with photoz inside the redshift bin, but some of the samplings falls out of the redshift bin $z \in [0.03,0.1]$ due to the photometric error incorporated in the PDF. 
The bottom panel illustrates another possible case.}
\label{fig:pdf}
\end{figure}

This sampling technique let us to produce $N_d = 1000$ data catalogs, termed sampled catalogs, that will act as fluctuations of our original data sample 
of $N_g = 5977$ blue galaxies. In figure~\ref{fig:pdf}, we can see two examples of those samplings for two specific galaxies together with the bin intervals used in this work. 
For some configurations this galaxy will be counted in the redshift bin and some others not. 
If the BAO signal is being biased by the configuration of our data, it should be dispersed by the fluctuations. 
On the other hand, if the BAO signal is present in most of the individual 2PACF analysis then it will appear in the average of the whole set of $N_d$ 2PACF.

\subsection{Determining \texorpdfstring{$\theta_{\rm BAO}$}{}} \label{subsec:recovering BAO}

After the complete 2PACF analyses shown in the previous sections, we shall compute our measurement $\theta_{\rm BAO}(z_{\rm eff})$, with its corresponding uncertainty and statistical significance.

\subsubsection{Parametric Model}\label{parametric}

To calculate the transversal BAO scale we used a parametric model proposed by~\citet{Sanchez2011} (see also~\citep{deCarvalho2020b}) 
that assumes a power law to describe the general shape of the correlation function plus a Gaussian function to fit the BAO bump 
\begin{equation} \label{eq:parametric model}
\omega(\theta) = A + B\theta^\gamma + 
Ce^{-(\theta - \theta_{\rm FIT})^2/2\sigma_{\rm FIT}^2} \,,
\end{equation}
where $A$, $B$, $\gamma$, $C$, $\theta_{\rm FIT}$, and $\sigma_{\rm FIT}$ 
are free parameters. 
The first three parameters incorporates the general behavior of the 2PACF also allowing it to have negative values, while the last three shows where and how much evident is the BAO signal within the data. 
The parameters $C$ and $\sigma_{\rm FIT}$ 
are related with the amplitude and dispersion of the BAO signature, while $\theta_{\rm FIT}$ indicates the angular scale of this signal. 
These parameters can be affected by the number of bins used in the 2PACF analysis, the redshift uncertainty of the galaxies, and the projection effect, that is, the dominant systematics producing the largest errors in our methodology.

\subsubsection{Projection Effect}\label{projection}

The BAO angular scale measurement would be directly obtained from the 2PACF for data in case of an infinitesimal thin redshift bin, $\Delta z \to 0$, because in such a case the projection effect would be negligible. 
However, this is not possible because 
we need a redshift bin large enough to have a significant amount of galaxies for analyses there. 
Distances along the line of sight are not taken into account when calculating the angular separation between pairs of galaxies with different redshifts, which generates a projection effect that shifts the expected scale for BAO \citep{Sanchez2011,carnero2011tracing, carvalho2016,deCarvalho2020}.

In order to achieve $\theta_{\rm BAO}$ with our parametric model we need to correct the projection effect in our $\theta_{\rm FIT}$. We can do this by calculating the expected angular BAO scale for an infinitesimal redshift bin, $\theta_E^0$, and for a bin of thickness $\delta z$, $\theta_E^{\delta z}$, 
\begin{equation}
\theta_{\rm BAO}(z) = \left[1 + \Delta \theta(z, \delta z)\right] \theta_{\rm FIT}(z)\,,
\end{equation}
where $\Delta \theta(z, \delta z) = (\theta_E^0 - \theta_E^{\delta z})/\theta_E^0$ is the percentage shift from the correct acoustic scale. To perform that calculation we need to assume a fiducial cosmology to calculate the 2PCF and then use equation~\eqref{eq:2PACF theory} to calculate the angular correlation function and derive those expected values. 

The projection effect can not only affect $\theta_{\rm BAO}$ by shifting $\theta_{\rm FIT}$, but also has the potential to affect the parameters $C$ and $\sigma_{\rm FIT}$ reducing the amplitude of the signal. 
To make this correction we had to assume a cosmological model, but \cite{Sanchez2011} shows that considering different models 
to calculate this shift only increases the error of the transversal BAO measurement by about $1\%$. 
Moreover, one can also introduce non-linearity when evaluating the angular shift $\Delta \theta$, and again this effect has a small impact when compared to the projection effect~\citep{deCarvalho2020b}. 
With these considerations for the sample in study, the percentage in angular shift due to the projection effect is $23\%$ or 
$\Delta \theta = 0.231$. 
To test the consistency of this result, we calculate it again but assuming two other fiducial cosmologies recently analysed by the DESI collaboration~\citep{DESI2025}, the flat-$\Lambda$CDM and $\omega_0 \omega_a$CDM, obtaining $\Delta \theta = 0.235$ and $0.228$, respectively. 
This confirms our result, and from now on we use 
$\Delta \theta = 0.23$ for correcting the $\theta_{\rm FIT}$ 
value due to the projection effect of our bin shell of thickness 
$\Delta z = 0.07$.

\subsection{Feldman-Kaiser-Peacock (FKP) weights}

The study of large-scale structure in galaxy distributions faces intrinsic problems to data catalogs. 
To minimize the sample variance and shot noise effects, the galaxy field has to be weighted. 
To do this, one assigns weights to each galaxy based on the average local density in the analysed region by using the Feldman-Kaiser-Peacock (FKP) weights 
\citep{Feldman94}. 
This is a scale-independent weighting that depends on redshift, 
\begin{equation}
w_{\text{FKP}}(z) = \frac{1}{1 + P_0\,n(z)} \, ,
\end{equation}
where $P_0$ is the amplitude of the power spectrum and $n(z)$ 
is the number density of galaxies in the data sample. 
We used $P_0 = 10000$ Mpc$^3\,h^{-3}$, the power amplitude 
relevant to the BAO signal $k \simeq 0.15$ 
Mpc $h^{-1}$~\citep{Eisenstein05,Beutler11,Carter18,deCarvalho2021}. 
Then, the effective redshift of our sample, $z_{\text{eff}}$, 
calculated with the FKP galaxy weights, 
$w_i \equiv w(z_i) = w_{\text{FKP}}(z_i)$, is obtained through 
\begin{equation}
z_{\text{eff}} = \frac{\sum^{n_g}_{i=1} w_i z_i}
{\sum^{n_g}_{i=1} w_i} \, ,
\end{equation}
where $n_g$ is the total number of galaxies in the data sample. 
Performing this calculation, we obtain the effective redshift 
$z_{\rm eff} = 0.075$ for our data sample.

\section{Results and discussions}\label{sec:results}

\begin{figure}
\centering
\includegraphics[width=0.9\linewidth]{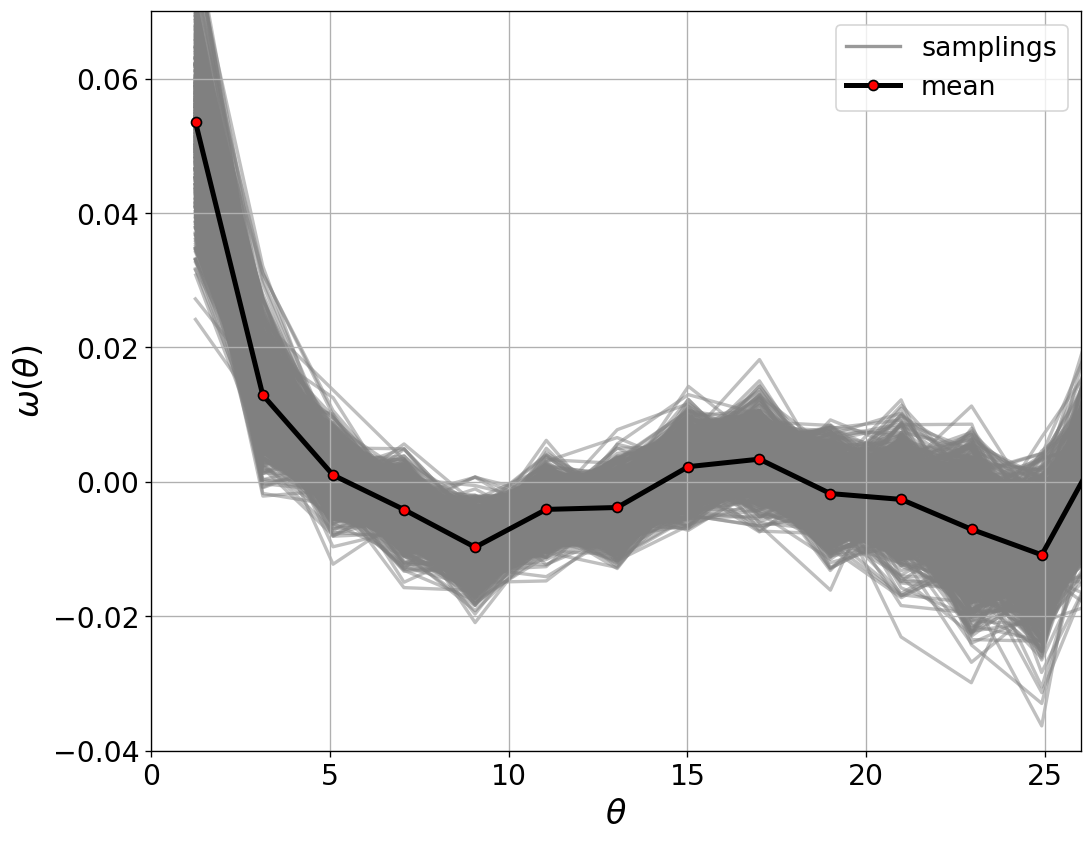} 
\includegraphics[width=0.9\linewidth]{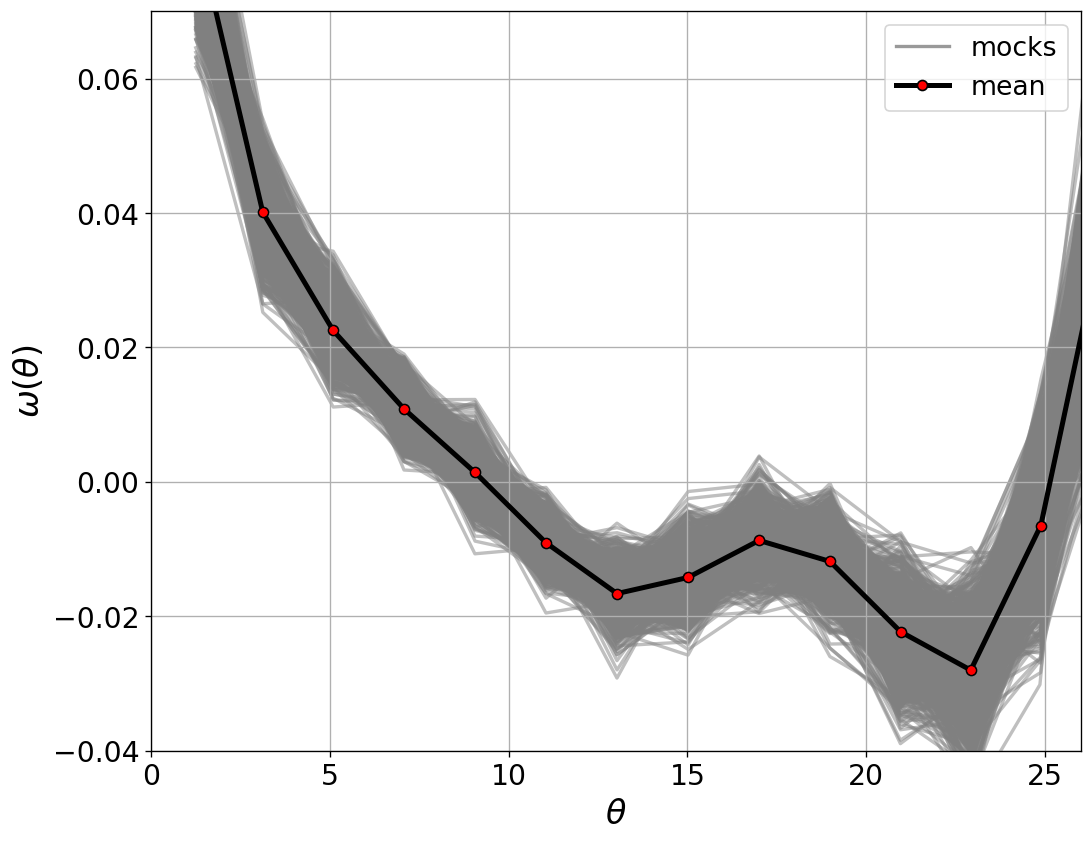}
\caption{\textbf{Upper panel}: 
The 2PACF analyses for the set of $N_d = 1000$ samplings generated from our original data catalog of $N_g = 5977$ photo-z blue galaxies; the continuous line represents the average of the set of correlation functions calculated for each sampled data. 
\textbf{Bottom panel}: Similar analysis to that performed in the upper plot, 
but this time considering the set of $N_m = 1000$ mocks.} 
\label{fig:2PACF}
\end{figure}

\begin{figure}
\centering
\includegraphics[width=1\linewidth]{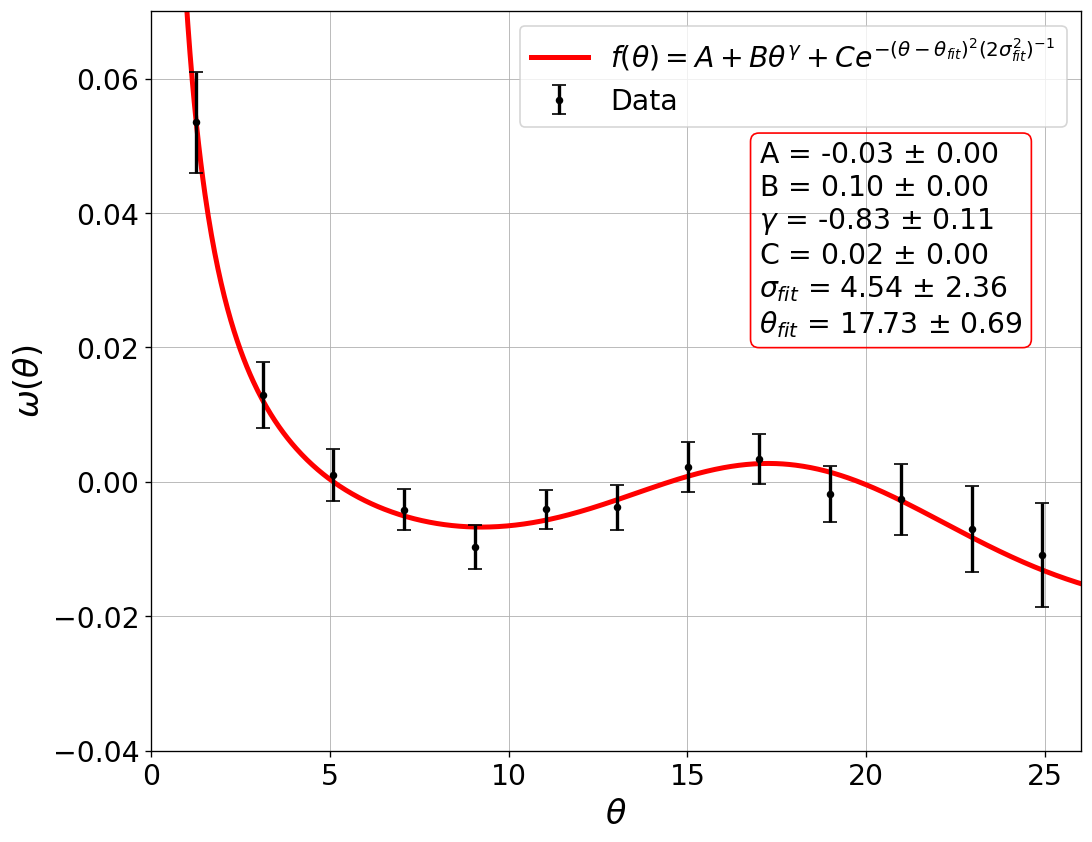}
\caption{
Using the data points corresponding to the mean 2PACF, shown in the upper panel of figure~\ref{fig:2PACF}, plus the bin errors calculated via the covariance matrix, computed in section~\ref{covariance-matrix} 
and illustrated in figure~\ref{fig:correlation-matrix}, 
we obtain the final 2PACF result shown in this plot. 
With this data bins, and their errors, we apply the best-fit 
procedure described in section~\ref{subsec:recovering BAO}, 
using equation~\eqref{eq:parametric model}, obtaining 
the best-fit parameters shown in the frame of this plot.
}
\label{fig:2PACF-final}
\end{figure}

The main results of our work are shown in figures~\ref{fig:2PACF} 
and~\ref{fig:2PACF-final}, where we display our analyses 
of the angle correlations between pairs of blue galaxies using the 2PACF. 
This study of the 2PACF was done considering a set of $N_d = 1000$ data catalogs produced by our sampling methodology, described in 
section~\ref{sampling-catalogs}. 

where some degree of dispersion is noticed around the mean curve, 
the continuous line in the upper panel of figure~\ref{fig:2PACF}, nevertheless the 2PACF clearly reveal the BAO signature as an excess of angular correlations around $17^{\circ}$. 

Additionally, in the lower panel of figure~\ref{fig:2PACF}
we show the analyses done considering the set of $N_m = 1000$ mock catalogs, originally produced to compute the covariance matrix for this measurement, but we found it important to use them for a robustness test. 
In fact, these analyses confirms the transversal BAO signature 
shown in the upper panel of figure~\ref{fig:2PACF}, obtained 
by sampling $N_d = 1000$ times our original data sample of 
photo-z blue galaxies.

%

The next step to find a BAO measurement is to perform the 
best-fit procedure, described in section~\ref{parametric}, 
on the 2PACF. 
To this end, we use the data points corresponding to the mean 2PACF, shown in the upper panel of figure~\ref{fig:2PACF}, plus the bin errors calculated via 
the covariance matrix, computed in section~\ref{covariance-matrix} 
and illustrated in figure~\ref{fig:correlation-matrix}, 
to obtain the final 2PACF result, which is shown in figure~\ref{fig:2PACF-final}. 
With this data bins with their corresponding errors we apply the best-fit 
procedure described in section~\ref{subsec:recovering BAO}, 
using equation~\eqref{eq:parametric model}, 
therefore obtaining the best-fit parameters shown in the 
frame of figure~\ref{fig:2PACF-final}, where 
\begin{eqnarray}
\theta_{\rm FIT} &=& 17.73^{\circ} \pm 0.69^{\circ} \,. 
\end{eqnarray}
The final step to obtain the transversal BAO measurement from the 2PACF is to include the angular shift due to the projection effect explained in section~\ref{projection}. 
In fact, due to the finiteness of the redshift bin, this effect shifts the angular scale $\theta_{\rm FIT}$ to $\theta_{\rm BAO}$ by a quantity that weakly depends on the cosmological model, but it strongly depends on the thick of the redshift bin. 
Accordingly, the result of our analysis is a percentage shift of 
$\Delta \theta = 0.23$, which means that the scale $\theta_{\rm FIT}$ 
suffers a $23\%$ shift; therefore, our transversal BAO measurement is 
\begin{equation}
\theta_{\textrm{BAO}} = 21.81^{\circ} \pm 0.85^{\circ} \,,
\end{equation}
at the effective redshift $z_{\rm eff} = 0.075$.

\begin{figure}
\centering
\includegraphics[width=1.1\linewidth]{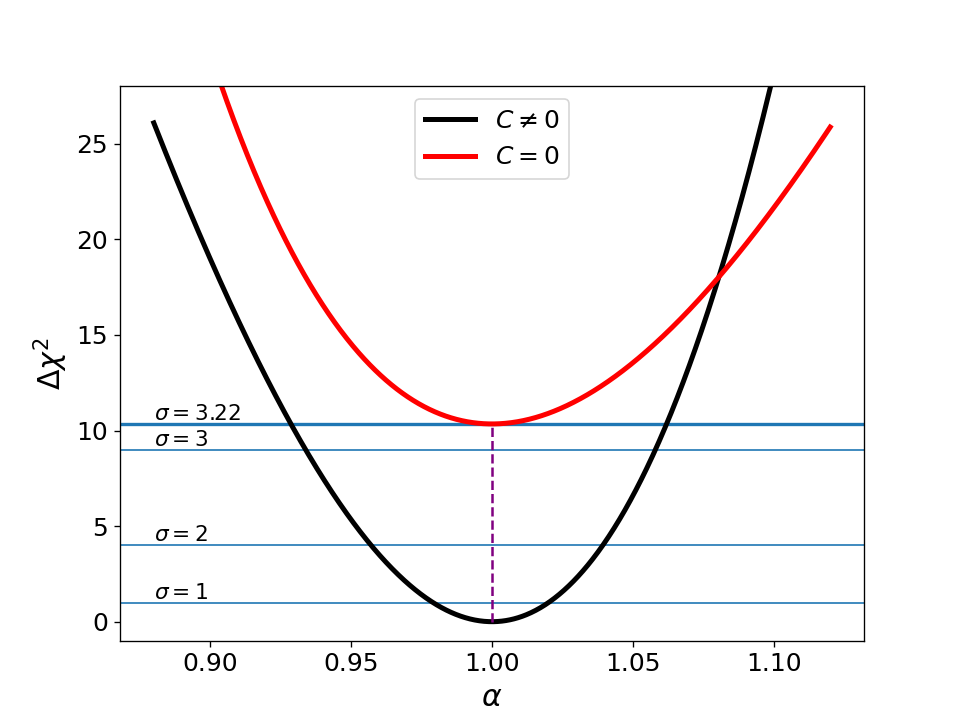}
\caption{Study of the statistical significance of our 2PACF measurement. 
Our analysis shows that our measurement of the transversal BAO, 
$\theta_{\textrm{BAO}} = 21.81^{\circ} \pm 0.85^{\circ}$ 
at $z_{\rm eff} = 0.075$, attains 3.22 $\sigma$ of confidence level. The red curve was made using only the power law terms of the parametric model given by equation \eqref{eq:parametric model}, whereas the black curve includes the Gaussian term.} 
\label{fig:stat_significance}
\end{figure}

Our final analysis concerns the statistical significance of this BAO measurement. This was obtained following \cite{deCarvalho2018} with a method based on the minimum $\chi^2$ as a function of the scale dilation parameter $\alpha$ for two cases: one using the parametric model \eqref{eq:parametric model}, but only with the power law terms $(C = 0)$, and the other case including the Gaussian term $(C \neq 0)$. We perform this calculation and find that our transversal BAO measurement 
attains the confidence level of 3.22\,$\sigma$, as illustrated in figure~\ref{fig:stat_significance}.

This angular scale measurement, $\theta_{BAO}$, 
works as a standard ruler if one consider from the literature 
a value for the sound horizon scale, $r_s$, because it is related to the diameter angular distance $D_A$,
\begin{equation*}
D_A(z) = \frac{r_s}{(1+z)\,\theta_{BAO}(z)} \,. 
\end{equation*}
Assuming, e.g., a value for 
$r_s = 99.08 \pm 0.18$ Mpc/h from~\citet{planck18}, 
one obtains 
$D_A(z_{\rm eff}) = 242.13 \pm 9.45$ 
Mpc/$h$, at $z_{\rm eff} = 0.075$.


\section{Conclusions and Final remarks}\label{sec:conclusion}

BAO, a physical process of the primordial universe, left a geometric signature in the observed distribution of matter structures. 
Although such signal can not be observed directly but statistically, 20 years of mapping the universe and its components were enough to achieve a high degree of confidence regarding this phenomenon. 
This confidence is supported not only by the increasing precision in the measurements of the sound horizon, and the related distance scales, but above all because it was detected in several cosmic tracers at different epochs of the universe evolution. 
Many BAO analyses have being done using catalogs with spectroscopic redshift data. 
However, photometric surveys are also currently available, and we find it interesting --and challenging-- the possibility to perform BAO measurements with these type of data.

For this enterprise we have developed an approach that uses the photometric redshift information of each blue galaxy to perform samplings of the original data generating a large set of 
$N_d=1000$ sampled catalogs. 
We then used the estimator Landy-Szalay to study the angular correlations of the blue galaxies in each one of these catalogs; 
then we concentrate in the analysis of the average of this set of $1000$ 2PACF, together with bin to bin uncertainties that were obtained calculating the covariance matrix from a set of previously produced mock catalogs. 
The statistical analyses of this 2PACF-average returns a best-fit angular BAO scale, $\theta_{\rm FIT}$ as observed in figure~\ref{fig:2PACF-final}, which one knows is 
affected by the projection effect and needs an angular shift correction~\citep{deCarvalho2020b,Sanchez2011}. 
After this procedure, one therefore obtain the transversal BAO scale $\theta_{\rm BAO}$ of the sample in analysis.

We have performed consistency tests along the analyses done, 
and also assessed the impact of photometric uncertainties on 
the BAO signal found recalculating the transversal BAO signature 
using simulated mock catalogs, as shown in the bottom panel of figure~\ref{fig:2PACF}.

Ultimately, the analysis of the S-PLUS blue galaxies in the 
redshift shell $z \in [0.03,0.1]$, resulted in the transversal 
BAO measurement: 
$\theta_{\textrm{BAO}} = 21.81^{\circ} \pm 0.85^{\circ}$, 
at the effective redshift $z_{\rm eff} = 0.075$. 
Finally, we also determine the statistical significance of 
this result. 
Our analysis, displayed in figure~\ref{fig:stat_significance}, 
shows that our measurement of the transversal BAO has 
3.22\,$\sigma$ of confidence level.
This transversal BAO measurement provides the angular diameter distance 
measurement $D_A(z_{\rm eff}) = 242.13 \pm 9.45$ Mpc/$h$, 
at $z_{\rm eff} = 0.075$.





\acknowledgments
UR thanks the Brazilian National Research Council (CNPq) for the financial support. CRB acknowledges the financial support from CNPq (316072/2021-4) and from
FAPERJ (grants 201.456/2022 and 210.330/2022) and
the FINEP contract 01.22.0505.00 (ref. 1891/22).  The
authors made use of Sci-Mind servers machines developed by the CBPF AI LAB team and would like to thank P. Russano  and M. Portes de
Albuquerque for all the support in infrastructure matters. FA thanks to Funda\c{c}\~{a}o Carlos Chagas Filho de Amparo \`{a} Pesquisa do Estado do Rio de Janeiro (FAPERJ), Processo SEI-260003/001221/2025, for the financial support. CF thanks Coordenação de Aperfeiçoamento de Pessoal de Nível Superior (CAPES) for the financial support.
AB acknowledges a CNPq fellowship. AC acknowledges the Funda\c{c}\~{a}o Carlos Chagas Filho de Amparo \`{a} Pesquisa do Estado do Rio de Janeiro (FAPERJ) grant E26/202.607/2022 e 210.371/2022(270993) and
the Brazilian National Research Council (CNPq).

\appendix
\section{Robustness test of the BAO measurement}
\label{appendixA}

\begin{figure}
\centering
\includegraphics[width=1\linewidth]{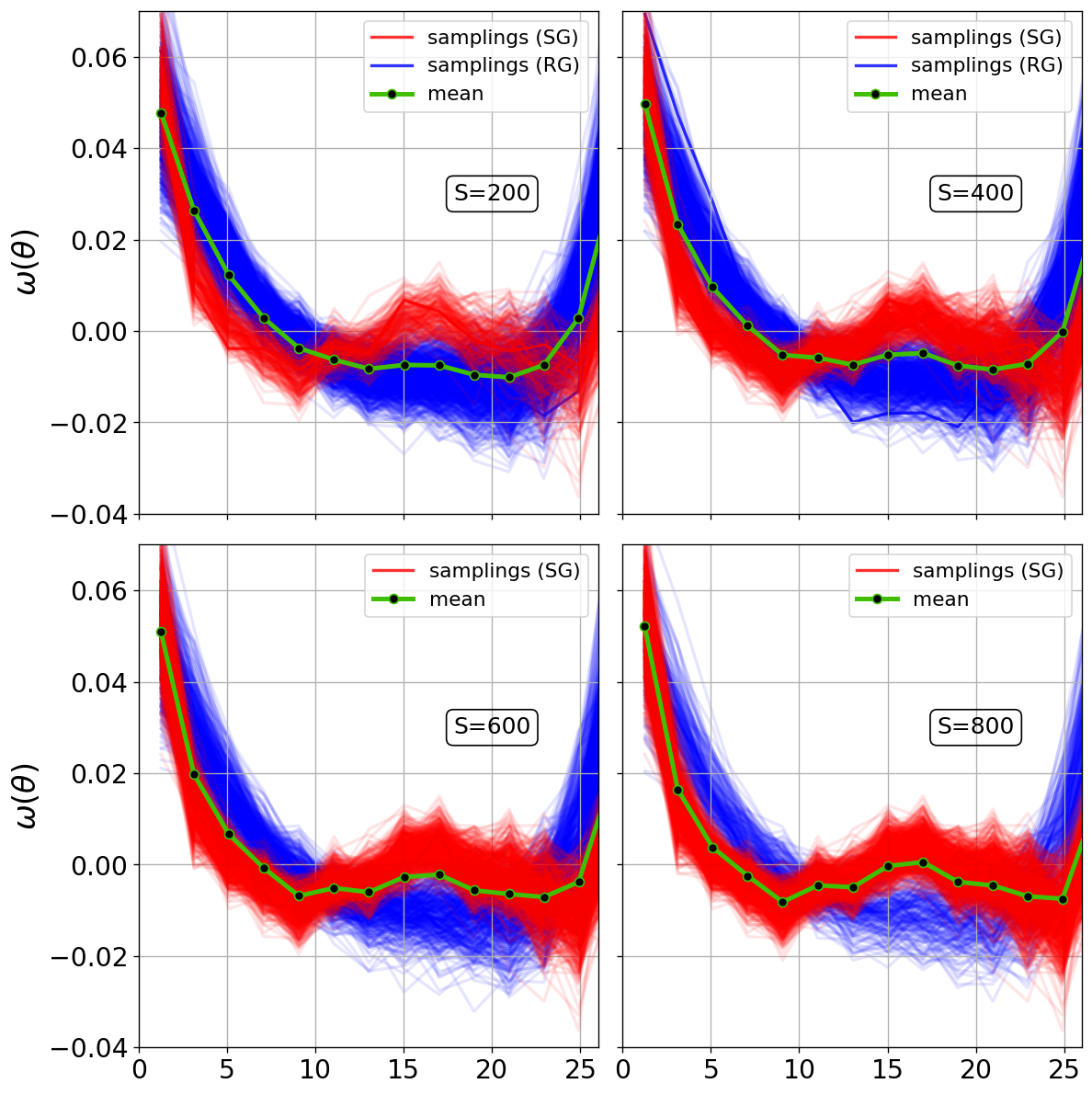}
\caption{These plots contain the 2PACF analyses considering 
the two sub-sets SG and RG defined in the Appendix~\ref{appendixA}. 
The cases in study consider S $= 200, 400, 600, 800$, that is, 
when averaged 2PACF (dots with a continuous line) contains 20\%, 40\%, 60\%, and 80\% of the original set of $1000$ catalogs of sampled data, respectively. 
%
}
\label{fig:sistematics_S}
\end{figure}

In this appendix, we test the robustness of our transversal BAO measurement, 
regarding a possible bias caused by the sampling approach of the photometric 
errors of blue galaxies. 
Firstly, we take the 1000 catalogues generated using the sampling 
methodology presented in section~\ref{sampling-catalogs}. 
We divide this set in two sub-sets: 
one with S $< 1000$ catalogs, termed the Signal Group, 
the other with $1000 - $S catalogs, termed the Randomized Group, 
specifically, 
\begin{itemize}
\item the Signal Group (SG): consists of the first S catalogs with 
their original angular coordinates; 
we have considered four cases: S $= 200, 400, 600, 800$. 
\item the Randomized Group (RG): 
consists of the rest of catalogs, i.e., $1000 - $S catalogs, where we randomize the angular positions of the galaxies, and the new coordinates are chosen according to normal distribution centred in the original coordinates 
%
\begin{subequations}
\begin{align*}
\textrm{RA}_{new} &= \textrm{RA}  + \mathcal{N}(0, \sigma) \,, \\
\textrm{DEC}_{new} &= \textrm{DEC} + \mathcal{N}(0,\sigma) \,, 
\end{align*}
\end{subequations}
where $\mathcal{N}(0,\sigma)$ is a random variable with a normal distribution centred at the original coordinate, with standard deviation $\sigma$. 
\end{itemize}
The analyses shown in figure~\ref{fig:sistematics_S} for the four 
cases of S were performed adopting $\sigma = 3$, a value large enough to destroy the BAO signal in the sub-set 
RG. 
In this figure we displayed the 2PACF for each catalogue, 
distinguishing with blue or red colour the 2PACF from each subset, 
then we averaged the whole set of 1000 2PACF. 
The result is clear, the BAO signature is present in the original set containing the sampled data and is washed out in the randomized catalogs.


\begin{figure}
\centering
\includegraphics[width=1\linewidth]{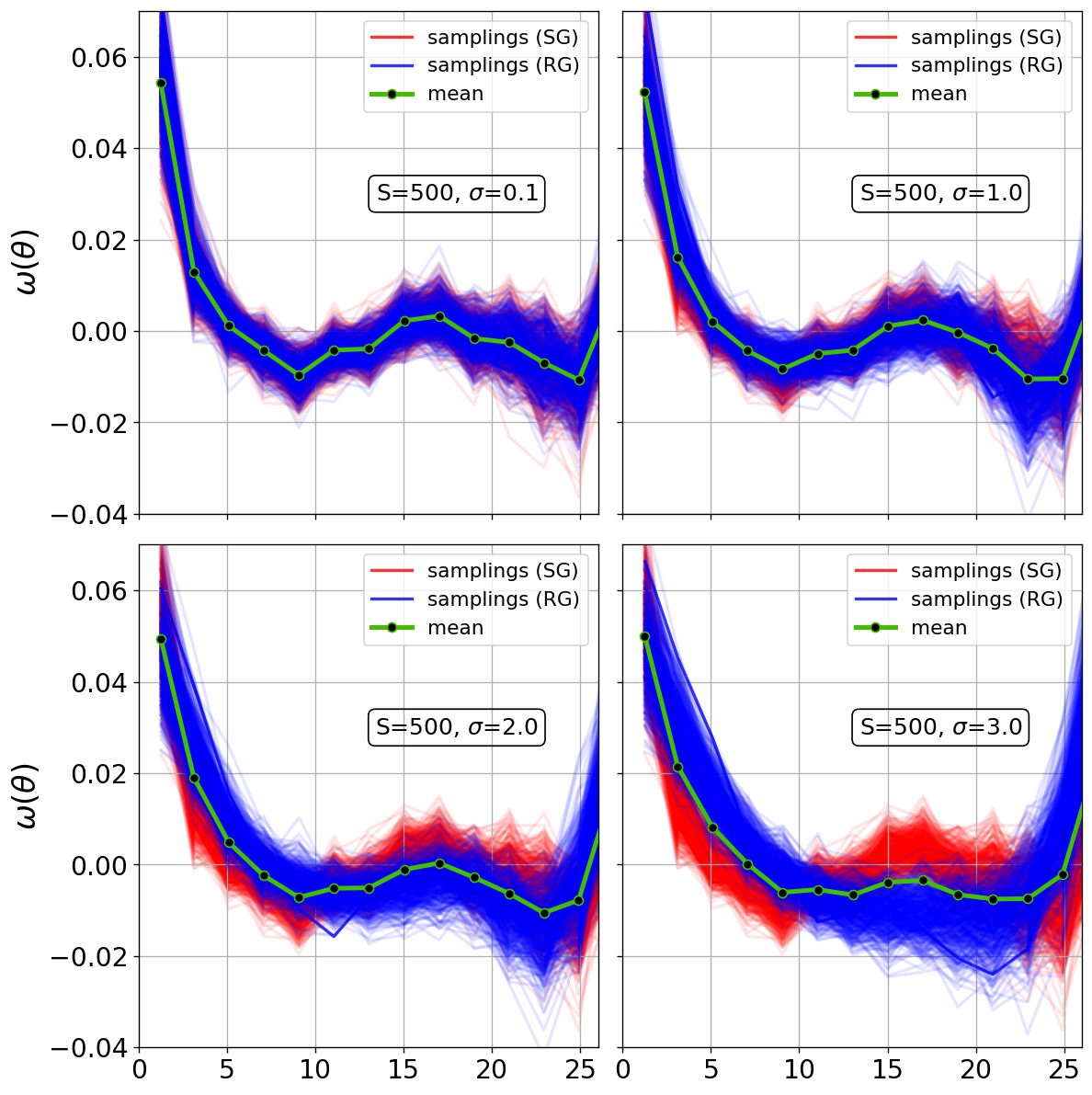}
\caption{These plots complement the analyses shown in the 
figure~\ref{fig:sistematics_S}. 
We analyse the case S $ = 500$ considering different values for the parameter $\sigma$, which quantifies the degree of randomization, 
in the sense: the larger $\sigma$, the greater the angular displacement of the galaxy position in relation to its original coordinates.
}
\label{fig:sistematics_sigma}
\end{figure}

For completeness, we also perform analyses considering different values for the standard deviation $\sigma$. 
In figure~\ref{fig:sistematics_sigma} we set $S = 500$ and gradually increase the value of $\sigma$ from 0.1 up to 3. 
As observed there, slight deviations from the original coordinates 
in the randomized catalogs do not destroy the BAO signal in the 
RG, as expected.

\bibliographystyle{sn-basic} 

\bibliography{sn-bibliography}

\end{document}